\documentclass[aps,amssymb,amsfonts,twocolumn,prl,floatfix,showpacs,superscriptaddress,10pt]{revtex4-1}

\usepackage{graphicx}
\usepackage{bm}
\usepackage{amssymb}
\usepackage{amsmath}
\usepackage{epsfig}
\usepackage[T1]{fontenc}
\usepackage{appendix}
\usepackage{color}
\usepackage{enumerate}
\usepackage{aeguill}
\usepackage{subfigure}

\graphicspath{{figs/}}

\begin{document}

\title{Laser filamentation as a new phase transition universality class}
\author{W. Ettoumi}
\email{wahb.ettoumi@unige.ch}
\affiliation{Universit\'e de Gen\`eve, GAP-Biophotonics, Chemin de Pinchat 22, CH-1211 Geneva 4, Switzerland}

\author{J. Kasparian}
\affiliation{Universit\'e de Gen\`eve, GAP-Non-linear, Chemin de Pinchat 22, CH-1211 Geneva 4, Switzerland}

\author{J.-P. Wolf}
\affiliation{Universit\'e de Gen\`eve, GAP-Biophotonics, Chemin de Pinchat 22, CH-1211 Geneva 4, Switzerland}

\date{\today}

\begin{abstract}
We show that the onset of laser multiple filamentation can be described as a critical phenomenon that we characterize both experimentally and numerically by measuring a set of seven critical exponents. This phase transition deviates from any existing universality class, and offers a unique perspective of conducting two-dimensional experiments of statistical physics at a human scale.
\end{abstract}


\maketitle
The extensive study of phase transitions has triggered off substantial experimental and theoretical achievements. Critical phenomena can be sorted out into universality classes, each one being characterized by a set of critical exponents. However, due to the intrinsic experimental difficulties arising when one needs to tweak control parameters, most of the theoretical predictions have only been tested numerically, leaving only a few experimental verifications of generic models~\cite{Back1995XPIsing,Takeuchi2009DP} available to date.

Population inversion at the core of laser phenomena has been the first example of an out-of-equilibrium second-order phase transition~\cite{Graham1970PhaseTr}. Since then, as the available optical power gradually increased throughout the years, laser physics opened the way to the field of non-linear optics, bringing upfront the study of solitons~\cite{Novoa2010a}, optical vortices~\cite{Curtis2003Vortices,Vuong2006OpticalVortices}, or filamentation~\cite{Braun1995,Moll2003SF,Kasparian2003}.

The latter phenomenon describes a self-guided propagation regime typical of high power lasers, and is initiated when the power exceeds a critical value~$P_\textrm{cr}$, allowing the Kerr self-focusing phenomenon to overcome diffraction. The self-focusing laser pulse then reaches intensities allowing medium ionization. The resulting plasma as well as other higher-order effects in turn contribute to the laser defocusing, therefore initiating a dynamic balance between these two competing effects.

When the input power largely exceeds $P_\textrm{cr}$, the beam breaks up into multiple cells that self-focus individually. This process is driven by optical turbulence~\cite{Kandidov1999,Hosseini2004a,Moloney1999Turbulence}, in which transverse modulational instability seeded by the irregularities of the laser intensity profile as well as the inhomogeneities of the atmospheric refractive index results in the local nucleation of filaments across the beam profile. The multiple filamentation pattern steadily builds up as the pulse propagates. Progressively, each individual filament reaches its clamping intensity by absorbing energy from the surrounding photon bath, that gets depleted. The beam therefore evolves from an initially noisy but homogeneous, fully-connected fluence state towards a clustered transverse section, with isolated plasma filaments surrounded by independent islands of moderate fluence.

This process results in typical patterns featuring as much as~1000 filaments, that grow at the expense of the photon bath intensity. Their spatial concentration is limited by lateral interactions, which prevent them from packing closer to one another beyond a density of several cm$^{-2}$~\cite{Hosseini2004,Ren2000,Henin2010a}.

Based on such experimental multiple filamentation patterns as well as numerical simulations, we show that during the beam propagation, a sharp transition occurs as the emergence of multiple filaments structures the beam into isolated high-fluence clusters separated by moderate-fluence areas. This evolution constitutes a phase transition, that we characterize by a set of seven critical exponents. Their values verify a hyperscaling universal relationship predicted by renormalization group arguments~\cite{Stanley1999RG}, thus confirming the existence of a critical phenomenon. However, in spite of the similarity with percolation patterns, the critical exponents deviate from those of the corresponding universality class, as well as from any other previously characterized system. The multiple filamentation onset therefore exhibits a new and experimentally accessible universality class, exempt from the usual drawbacks of sample preparation or issues on the control parameters. Our findings may also pave the way towards an original theoretical framework for investigating the effect of the geometry of multiple filamentation on applications like laser-induced condensation \cite{Henin2011} or the triggering of lightning \cite{Kasparian2008d}, where the mechanism at a microscopic scale is strongly impacted by connectivity within the laser beam.

\bigskip

\begin{figure*}[htbp]
\begin{center}
\includegraphics[width=2.00\columnwidth]{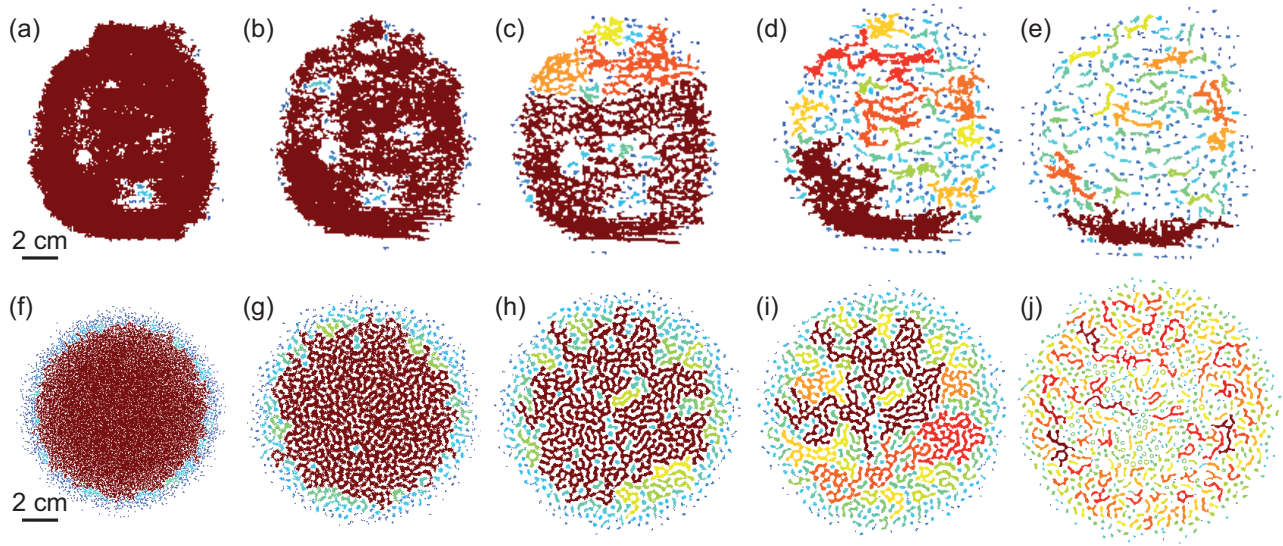}

\end{center}
\caption{Formation of a structured disconnected filamentation pattern from an initially smooth one in a $100$~TW beam. Experimental record (a-e) and numerical simulation (f-j) evolution of the fluence profile, after propagating in air for (a) $1.25$~m; (b) $2.5$~m; (c) $5$~m; (d) $10$~m, and (e) $15$~m. Numerical simulations are shown here for (f) $1$~m; (g) $8$~m; (h) $10.6$~m; (i) $11.7$~m and (j) $14.4$~m. The fluence has been thresholded at the same value in all pictures, corresponding to $0.8\mathrm{TW}/\mathrm{cm}^2$ for numerical simulations. Clusters are colored according to their relative size in each image.}
\label{fig:maps}
\end{figure*}

Experimental fluence patterns~\cite{Henin2010a} from a $100$~TW beam recorded on photographic paper over $15$~m of propagation (Figure~\ref{fig:maps}a-e) illustrate the formation of a multiple filamentation pattern. From a single connected cluster (panel a), the depletion of the photon bath splits the beam into several distinct clusters (panels c-e), colored by size in Figure~\ref{fig:maps}. Details of the image processing can be found in the Supplementary Material.

\begin{figure}
	\centering
		\includegraphics[width=1.00\columnwidth]{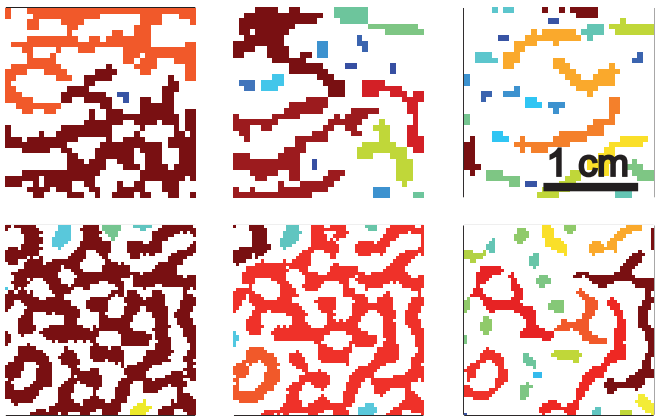}
			\caption{Close-up on the (a-c) experimental and (d-f) numerical multiple filamentation patterns of Figure~\ref{fig:maps}, at propagation distances (a) $5$~m; (b) $10$~m, and (c) $15$~m for the experimental data, and (d) $10.6$~m; (e) $11.7$~m and (f) $14.4$~m for their numerical counterparts. Clusters corresponding to laser filaments survive all along the propagation, even when the filaments begin to vanish at long distances.}
	\label{fig:lattices}
\end{figure}

A closer look at the cluster structure (Figure~\ref{fig:lattices}a-c) shows that, during the propagation, the ``strings'' connecting the filaments vanish, while small-size ($\approx$ 2~mm$^2$) clusters survive around individual plasma filaments.

The transition from a single cluster connecting the opposite sides of the beam profile, to a disconnected, multi-clustered pattern, appears to occur sharply between $z=5$ and $10$~m, reminding us of a phase transition in a percolation experiment.

In order to overcome the intrinsically limited amount of experimental data and their finite size, we sampled square lattices of different sizes from each image using a coarse-graining procedure. In each map, we measured the filling factor~$p$, defined as the ratio between the number of occupied sites and their total number, the spatial correlation length~$\xi$, as well as the mean cluster surface~$S$. In the latter measurement, following the usual practice of percolation analysis~\cite{Fortunato2001Perco}, the percolating cluster is excluded.

\begin{figure}
	\centering
		\includegraphics[width=1.0\columnwidth]{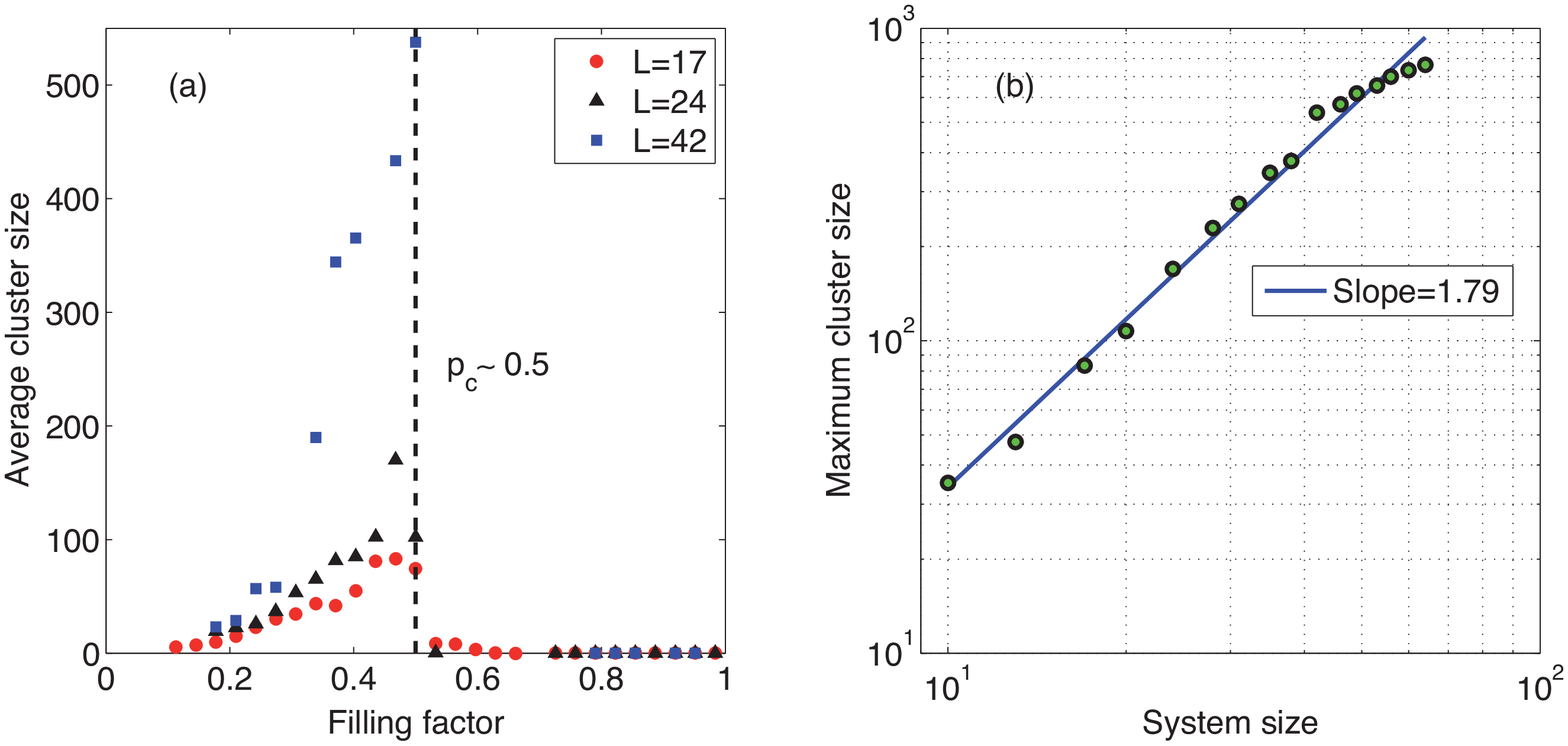}
	\caption{(a) Experimental mean cluster size (surface) as a function of the lattice filling factor~$p$. (b) Maximum cluster size with respect to the system size $L$. We disregarded the percolating cluster appearing when $p$ exceeds $p_\mathrm{c}=0.5$ from the surface calculation.}
	\label{fig:FSS_xp}
\end{figure}

As displayed in Figure~\ref{fig:FSS_xp}a, the mean cluster size exhibits a maximum value when $p$ reaches close to a critical value $p_\mathrm{c}\approx0.5$.
Close to such criticality, for spatially infinite systems, the phase transition theory predicts a divergence of the spatial correlation length $\xi$ as well as the mean cluster size $S$, defining two critical exponents $\nu$ and~$\gamma$:
\begin{eqnarray}
\xi &\sim& \left|p-p_\mathrm{c}\right|^{-\nu}. \label{eqn:scalingXi} \\
S &\sim& \left|p-p_\mathrm{c}\right|^{-\gamma}, \label{eqn:scalingS}
\end{eqnarray}
In a finite system, $\xi$ is bounded to the width $L$ of the sampled squares. Near criticality, it can therefore be assumed to lie close to this value, so that the previous scalings rewrite as:
\begin{eqnarray}
\xi &\sim& L \sim \left|p-p_\mathrm{c}\right|^{-\nu} \label{eqn:scalingXi_finite} \\
S &\sim& L^{\frac{\gamma}{\nu}}
\label{eqn:scalingS_finite}
\end{eqnarray}
Based on this assumption, we fitted the~$\gamma/\nu = 1.79 \pm 0.1$ ratio from our samples (Figure~\ref{fig:FSS_xp}b). This value is independent within $\pm 0.1$ from the parameters of the coarse-graining procedure.

Our result displays multiple filamentation as a very promising practical experimental system to investigate phase transitions. Indeed, as the propagation distance $z$ increases, the filaments progressively emerge from the noise, and the filling factor $p$ monotonically and \emph{continuously} decreases as the intensity profile structures itself. In addition to being directly observable, this continuous and monotonic decrease also provides an easily and finely tunable control parameter for the said phase transition. Such property contrasts with many systems, where experimental constraints drastically limit the fine-tuning of the control parameters, and in many cases only allow measurement in a few conditions~\cite{Vasseur2012SpinGlass,Dahmani1985PhaseD,Zhang2010PhaseD}, at the cost of complex and costly sample preparation~\cite{Back1995XPIsing}.

Willing to fully characterize the phase transition at stake here, we turned to numerical simulations in order to supplement the experimental observations with a large amount of realizations and continuous measurements along the propagation distance.

\bigskip

We simulated the propagation of a flat-top beam, with 1~TW/cm$^2$ incident intensity. As the experimental data exhibits a high shot-to-shot reproducibility, we infer that the initial phase mask is mainly governed by the laser profile rather than the atmospheric turbulence. Therefore, we modeled this intrinsic perturbation by adding a $10\%$ white noise on amplitude and a $0.2$ radian amplitude noise on the complex phase onto the input profile, rather than using turbulence masks~\cite{Kandidov1999}. Both of these noises were made continuously differentiable at the numerical grid precision thanks to an interpolation method.

Comparing panels a-e and f-j of Figure~\ref{fig:maps}, as well as close-up panels a-c and d-f of Figure~\ref{fig:lattices}, respectively, shows that the model results exhibit major similarities with the experimental patterns, even if the detailed arrangement of the individual filaments differs to some extent. In particular, the emergence of connected filament clusters is adequately reproduced, as well as their size distribution, its evolution during propagation, and the subsequent splitting of the initial clusters into smaller ones of $2~\mathrm{mm}^2$ around the location of each filament.

\begin{figure}[t]
	\centering
		\includegraphics[width=1.00\columnwidth]{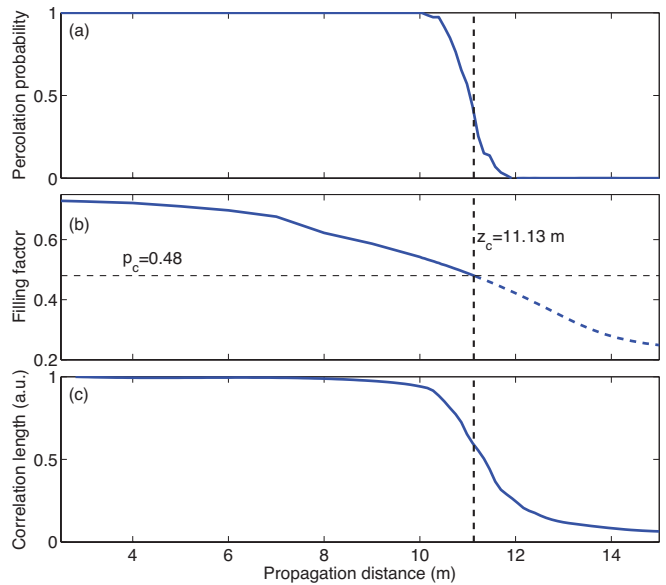}
	\caption{Evolution of (a) the percolation probability, (b) the filling factor $p$, and (c) the correlation length as a function of the propagation distance $z$, defining the phase transition between the percolating and non-percolating states.}
	\label{fig:zp}
\end{figure}

The simulated patterns were processed the same way the experimental ones were. Figure~\ref{fig:zp} summarizes the results of this analysis, averaged over 92 independent noise realizations. As $z$ increases, the filling factor monotonically decreases, causing the breakdown the correlation length as well as the fragmentation of percolating fluence patterns into multiple sets of non-percolating variously sized clusters.

Besides the scaling laws on the average cluster size and correlation length (Equations (\ref{eqn:scalingXi}) and (\ref{eqn:scalingS})), two supplementary ones are accessible (see details in Supplementary Material). They concern the cluster strength $P_\infty$ (i.e., the probability that a site belongs to the percolating cluster), as well as the the typical maximum cluster size $s_\mathrm{max}$. As expected for a critical phenomenon:
\begin{eqnarray}
P_\infty &\sim& (p-p_\mathrm{c})^{\beta},\quad p\geq p_\mathrm{c}\label{eqn:scaling_Pinf}\\
s_\mathrm{max} &\sim& |p-p_\mathrm{c}|^{-\sigma}. \label{eqn:scaling_smax}
\end{eqnarray}

As summarized in Table~\ref{tab:cr_exponents}, the more detailed statistical analysis brought by the numerical simulations allows one to measure these critical exponents. Three more exponents were deduced from scaling relations, as detailed in the Supplementary Material. The measurement of seven critical exponents in a single experiment is exceptional since one often only has access to a few ones~\cite{GinelliDP2003,Tephany1997Perco}. Up to date only K.~Takeuchi~\cite{Takeuchi2009DP} achieved such a comprehensive characterization, for the specific case of directed percolation.
\begin{table}[htbp]
\begin{tabular}{cccc}
Critical & Filamentation & Filamentation & 2D uncorrelated \\
exponent & experiment & simulation & percolation \\
\hline
\hline
$\beta$ & - & $0.13\pm0.02$ & $5/36\approx0.14$ \\
$\gamma$ & - & $3.0\pm0.3$ & $43/18\approx2.34$ \\
$\nu$ & - & $1.6\pm0.1$ & $4/3\approx1.33$ \\
$\gamma/\nu$&$1.79\pm0.1$ & $1.91\pm0.05$ & $43/24\approx1.79$ \\
$\sigma$ & - & $0.33\pm0.02$ & $39/91\approx0.39$\\
\hline
$\alpha$ & - & $-1.2^\ast\pm0.2$ & $-2/3\approx-0.67$ \\
$\eta$ & - & $0.09^\ast\pm0.29$ & $5/24\approx0.21$ \\
$\delta$ & - & $24.9^\ast\pm6.5$ & $91/5=18.2$ \\
\hline
$p_\mathrm{c}$& $0.5\pm0.1$ & $0.48\pm0.01$ & $0.5927$\\
\hline
\end{tabular}
\caption{Critical exponents measured from multiple filamentation experiments and simulations, and expected values for two-dimensional uncorrelated percolation. The asterisks indicate that a value has been derived from a hyperscaling relation, and not from a direct finite-size scaling measurement.}
\label{tab:cr_exponents}
\end{table}

Surprisingly, in contrast with the visual intuition from the similarity between percolation and multiple filamentation patterns (Figure~\ref{fig:maps}), these values demonstrate that the optical phenomenon does not belong to the same universality class as uncorrelated two-dimensional percolation.

\begin{figure}[htbp]
\begin{center}
	\includegraphics[width=1.00\columnwidth]{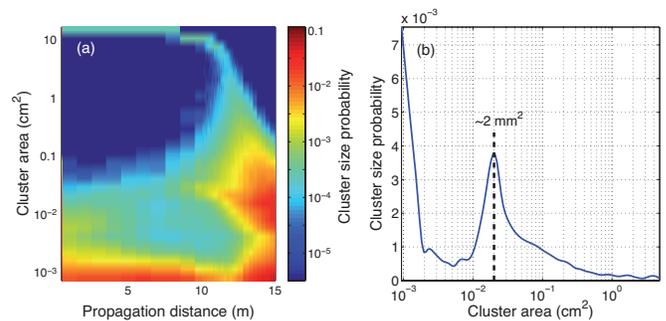}
\end{center}
\caption{(a) Evolution of the normalized cluster size distributions $n_s$ with respect to the propagation distance. A large percolating cluster is visible until approximately $z=10$m. Beyond, a mode emerges around $2~\mathrm{mm}^2$, corresponding to a robust size associated with individual filaments. (b) Normalized cluster size distribution for $z=11.7$m with respect to the cluster area in log scale. The typical size corresponds to the spatial extension of the photon bath immediately surrounding a plasma filament.}
\label{fig:clustersize}
\end{figure}

This discrepancy is illustrated at long distances by the survival of clusters of a typical size of $2$~mm$^2$ (also seen in the experiments), even when $p$ decreases down to $0.3$. These are clearly visible in blue in Figure~\ref{fig:lattices}, and appear as a ``topological defect'' in the cluster sizes distributions (Figure~\ref{fig:clustersize}).

Physically, this survival of clusters of well-defined size stems from the robustness of plasma filaments once they are initiated. Indeed, the dynamic balance at the root of filamentation can, e.g. withstand turbulence strengths 5 orders of magnitude beyond natural atmospheric values~\cite{Mechain2005Adverse}. This feature clearly diverges from uncorrelated percolation expectations, where the cluster size distributions obey a scaling law~$n_p(s)=s^{-\tau} g_{\pm}\left[(p-p_\mathrm{c})s^\sigma\right]$ close to criticality. Obviously, this equation cannot fit the size distribution corresponding to multiple filamentation (Figure {\ref{fig:clustersize}}).

It therefore appears that the emergence of multiple filamentation constitutes a phase transition. The reliability of our results is assessed by checking the universal scaling relation $\nu d = 2\beta + \gamma$, where $d = 2$ is the geometrical dimensionality of our system. In our case, we find $(2\beta + \gamma)/\nu=2.0\pm0.3$, ensuring that we indeed deal with a critical phenomenon. However, the set of seven critical exponents that we were able to determine from the numerical and experimental data matches none of known universality classes, suggesting that the patterns of laser multiple filamentation defines a new class by itself. 

Let us underline the fact that the critical exponents determined here are fundamentally \emph{equilibrium} quantities, therefore associated to a given \emph{equilibrium} situation at each considered propagation distance. If the patterns were to be considered as analogues of a classical spin system in the microcanonical ensemble, each image would then be an equilibrium configuration corresponding to a given temperature, controlled here by the propagation distance. In spite of the time evolution of the electric field, the consistency of our results indicates that the evolution of the filamentation pattern is quasi-static. Such a steady evolution can be related to the aforementioned filament robustness. 

It is unexpected to witness the success of equilibrium finite-size scaling techniques and their non-trivial predictions on patterns extracted from the integration of a differential equation as well as from real experimental material. The reason is twofold: firstly, the observed patterns arise from a modulational instability, but nothing \textit{a priori} suggests that non-trivial scaling laws would actually apply. Secondly, the variation of the filling factor with the propagation distance is intimately related to a time evolution, contradicting the standard picture of equilibrium statistical physics.

\bigskip

In conclusion, we have shown, based on both experiments and numerical simulations, that the emergence of multiple filamentation in a high-intensity, ultrashort laser beam is a critical phenomenon. This opens the opportunity to realize two-dimensional percolation-like experiments at a human scale and relate to future theoretical developments. Moreover, the values of the associated seven critical exponents evidences that this phenomenon belongs to a previously unidentified universality class.

Beyond providing a first prototype for this universality class, the propagation of ultrashort laser pulses also offers a very rich experimental and numerical benchmark, since the control parameter, namely the filling factor $p$, can be continuously tuned and controlled from 0 to 1 via the propagation distance $z$.

We also expect the recently proposed atmospheric applications of filamentation to directly benefit from our findings. For instance, the triggering of lightning~\cite{Kasparian2008d} requires a high conductivity along the laser beam and therefore relies on connectivity between the multiple filaments, and the phase transition evidenced in the present work may have crucial impacts on the efficiency of these processes.

\begin{acknowledgments}
We acknowledge the Forschungszentrum Dresden-Rossendorf (FZD) contribution (groups of Pr.~R.~Sauerbrey and Pr.~U.~Schramm) to the experiments obtained with the DRACO facility. Fruitful discussions with R.~Vasseur, H.~Duminil-Copin and N.~Berti are also gratefully acknowledged. We acknowledge support from the European Research Council Advanced Grant ``Filatmo''.
\end{acknowledgments}

\section{Numerical simulations}

In order to achieve the numerical simulation of the multifilamentation phenomenon, we turned to a two-dimensional time-integrated model (2D+1)~\cite{Skupin2004}. Using a factorization ansatz for the real electric field, namely $\mathcal{E}=A(x,y,z) \times \chi(t)$, one obtains an effective equation for the field envelope $A$:

\begin{equation}
\partial_z A = \frac{\mathrm{i}}{2k_0} \Delta_\bot A + \left[\mathrm{i} \alpha k_0 n_2 |A|^2 - \mathrm{i} \gamma |A|^{2K} - \frac{\beta^{(K)}}{2\sqrt{K}}|A|^{2K-2}\right] A,
\label{eqn:prop_complete}
\end{equation}
where $\Delta_\perp$ is the transverse Laplacian, $k_0$ is the central wave-vector, $\alpha$ and $\gamma$ parameters modeling the delayed medium response, $n_2$ the Kerr non-linear refractive index, and $\beta^{(K)}$ the multiphoton absorption coefficient ($K$=8 in air). After rescaling the transverse dimensions by $1/\sqrt{2k_0} \left(\gamma/(\alpha k_0 n_2)^K\right)^{1/(2K-2)}$ and introducing the convenient set of dimensionless variables
\begin{eqnarray}
\eta&=&z(\gamma/(\alpha k_0 n_2)^K)^{-1/(K-1)},\\
\nu&=&\beta^{(K)}/(2\sqrt{K} \left(\alpha k_0 n_2 \gamma^{K-2}\right)^{-1/(K-1)},\\
\psi&=&A (\alpha k_0 n_2/\gamma)^{-1/(2K-2)},
\end{eqnarray}
eq.~(\ref{eqn:prop_complete}) rewrites under a compact generic form:
\begin{equation}
\mathrm{i} \partial_\eta \psi + \Delta_\bot \psi + |\psi|^2 \psi + F(|\psi|^2)\psi= 0,
\label{eqn:NLS_adim}
\end{equation}
where the underlying physics is described by the function~$F$
\begin{equation}
F(|\psi|^2)=-|\psi|^{2K} +\mathrm{i}\nu |\psi|^{2K-2}\label{eqn:models_plasma}.
\end{equation}

We performed the numerical integration of eq.~(\ref{eqn:NLS_adim}) using a split-step Fourier method and set an adaptative propagation step $\delta \eta$ proportional to $|\psi|_\mathrm{max}^{-1}$ in order to resolve the sharp intensity peaks. We assessed that the resolution of the numerical grid ($1.8~\mathrm{\mu m}$ per pixel in the transverse direction) was sufficient, by checking the conservation of the system hamiltonian when losses due to the ionization were neglected (i.e., setting $\nu=0$). More specifically, we monitored the conservation of the hamiltonian~$\mathcal{H}$:
\begin{equation}
\mathcal{H}=||\nabla_\bot \psi||^2_2-\frac{1}{2} ||\psi||^4_4+\frac{1}{K+1} ||\psi||^{2K+2}_{2K+2},
\end{equation}
where $||\cdot||_n=[\int |\cdot|^n]^{1/n}$ denotes the standard $L^n$ norm. Energy conservation was indeed ensured within less than $1\%$ error over the propagation distance.

Initial conditions were chosen as close as possible to the experimental situation. The flat-top beam was modelled by a radially symmetric fourth-order supergaussian, $\psi(r)=\psi_0 \mathrm{e}^{-(r/\sigma)^4}$. The amplitude $\psi_0$ was calculated from the initial power~$P_\mathrm{in}$ by
\begin{equation}
 \quad \psi_0= \sqrt{\frac{2^{1/4}P_\mathrm{in}}{\pi \sigma^2 \Gamma(5/4)}} \left(\frac{\alpha k_0 n_2}{\gamma}\right)^{-1/(2K-2)}.
\label{eqn:amplitude}
\end{equation}
To ensure a continuously differentiable intensity and phase profile, the noise was first calculated on a coarse mesh, then interpolated on our 4~fold larger numerical grid. Intensity profiles are stored over the whole propagation length of $15~\mathrm{m}$ for subsequent analysis.

\section{Recording of experimental patterns}

As described in detail earlier~\cite{Henin2010a}, experiments were performed using a titanium:sapphire chirped-pulse amplification providing $3$~J, $100$~TW pulses of $30$~fs duration, at a repetition rate of $10$~Hz and central wavelength of $800$~nm. The beam was launched collimated in air with a diameter of approximately $10$~cm, through a 6 mm thick fused silica window, the dispersion of which was pre-compensated by adequately adjusting the grating compressor of the laser system. The multiple filamentation of the beam was characterized by single-shot burns on the back side of photosensitive paper (Kodak Linagraph 1895). The images were then scanned into a grayscale colormap and processed as described in detail below.

\section{Image processing and analysis}

The beam fluence profiles, whether experimental or numerical, corresponding to a propagation distance $z$, was handled as a grayscale image. The fluence was first thresholded to a fixed value slightly below the initial fluence level of the initial beam ($0.8~\mathrm{TW}/\mathrm{cm}^2$ in the numerical simulations). 
We then coarse-grained the resulting black and white images on square lattices of $16\times16$ pixels.

Each bin was declared active or inactive (respectively filled or empty) depending on the proportion within the bin, of pixels above the fluence threshold from the original profile. We checked that the results were independent from the cell dimension as long as the coarse-graining was relevant.

In the case of the experimental data, where only one realization was available, a statistical analysis was made possible by excluding the regions where diffraction patterns are clearly visible at the bottom of the beam profile, and by randomly sampling square systems of different sizes over the spatial region of interest ($10\times10$ to $64\times64$).
In the case of numerical results, the large number of realizations obtained at each distance allowed us to focus, for each of them, on the central square region of approximately $6~\text{cm}\times6~\text{cm}$ in the $50$~TW case and $8~\text{cm}\times8~\text{cm}$ in the $100$~TW case, corresponding to the beam waist, on which we conducted all of our numerical measurements.

We define a cluster as a set of active sites connected through nearest-neighbor interaction. Following~\cite{Fortunato2001Perco}, we define percolation when a cluster spans the lattice in both the horizontal and the vertical directions, in order to ensure that one percolating cluster can exist at most. We calculate the percolation probability~$\Pi(z)$, at any distance~$z$, as the average number of runs which percolate at this distance, divided by the total number of runs. We also investigated the the maximum cluster size $s_\mathrm{max}$, the cluster strength $P_\infty$ (defined as the ratio between the largest cluster size and the system surface $L^2$), the filling factor $p$, the average cluster size~$S$, as well as the correlation length~$\xi$. The three latter observables are respectively defined as:
\begin{eqnarray}
p&=&\sum_{s=1}^{L^2} s n_s \\
S&=&\frac{1}{p} \sum_{s=1}^{L^2} s^2 n_s \\
\xi^2 &=& \frac{\sum_r r^2 g(r)}{\sum_r g(r)},
\end{eqnarray}
where $g(r)$ is the probability that two active sites separated by a distance~$r$ belong to the same cluster, i.e., are continuously connected by active sites. $n_s$ is the cluster size distribution, normalized by the system size $L$.
Note that these values can be calculated for any given propagation distance $z$, or, alternatively, for any filling factor~$p$. 

In the latter case, one has to take into account that, for each realization of the pulse propagation, the values of $p$ fluctuate for a given $z$. Therefore, in order to perform consistent ensemble averages, we weighted the observables by assuming that $p$ was distributed around its average $\bar{p}(z)$ following a Gaussian law of variance $\sigma^2(z)$ at each step, so that if one has $n$ samples for a given $z$, the average of any observable $\mathcal{O}$ reads
\begin{equation}
\left\langle \mathcal{O}(z) \right\rangle \propto \sum^{n}_{k=1} \mathcal{O}_k \mathrm{e}^{-(p_k-\bar{p}(z))^2/(2\sigma^2(z))},
\end{equation}
where the normalization factor has been omitted for the sake of clarity.

\section{Finite-size scaling}

Given the finite size of our samples, we determined the critical exponents $\beta$, $\gamma$, $\nu$ and~$\sigma$ governing the evolution of $P_\infty$, $S$, $\xi$, and $s_\mathrm{max}$, respectively, close to the phase transition, by using finite-size scaling techniques. Such techniques rely on the ansatz that close to the percolation threshold, the correlation length is of the same order of magnitude as the system size $\xi \sim L$.

\begin{figure}[t]
\begin{center}
 \includegraphics[width=1.0\columnwidth]{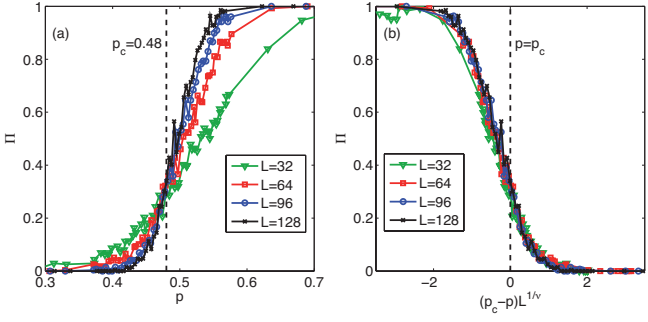} 
\end{center}
\caption{(a) Percolation cumulant~$\Pi$ with respect to the filling factor~$p$. The expected unique crossing point defines~$p_\mathrm{c}=0.48\pm0.01$. All the curves result from the average of $92$ independent runs. (b) Percolation cummulant $\Pi$ as a function of the rescaled variable $(p_\mathrm{c}-p)L^{1/\nu}$, yielding $\nu\approx 1.6\pm0.1$ by curve collapse.} 
\label{fig:FSS_S_th}
\end{figure}

The first step is to determine the phase transition threshold~$p_\mathrm{c}$. Let us for that purpose define the percolation cumulant~\cite{Fortunato2001Perco} as the fraction $\Pi$ of configurations featuring a percolating cluster. Since $p-p_\mathrm{c} \sim L^{-\frac{1}{\nu}}$ close to the percolation threshold, the scaling hypothesis allows one to write
\begin{equation}
\Pi\left(p\sim p_\mathrm{c},L\right)=g_\pi\left[L^{\frac{1}{\nu}} \left(p_\mathrm{c}-p\right)\right],
\end{equation}
where $g_\pi$ is a generic scaling function. Note that the definition of~$\Pi$ is only valid on finite lattices. It converges towards a step function at $p=p_\mathrm{c}$ when $L\to\infty$. Plotting~$\Pi$ with respect to $p$ for different system sizes exhibits a unique crossing point (Figure~\ref{fig:FSS_S_th}a), therefore yielding the value of $p_\mathrm{c}=0.48\pm0.01$. Plotting~$\Pi$ with respect to~$L^{\frac{1}{\nu}} \left(p_\mathrm{c}-p\right)$ allows one to obtain a first estimation of~$\nu$ by curve collapse. Figure~\ref{fig:FSS_S_th}b shows the determination of $\nu=1.6\pm0.2$.

\begin{figure}[t]
\begin{center}
 \includegraphics[width=1.0\columnwidth]{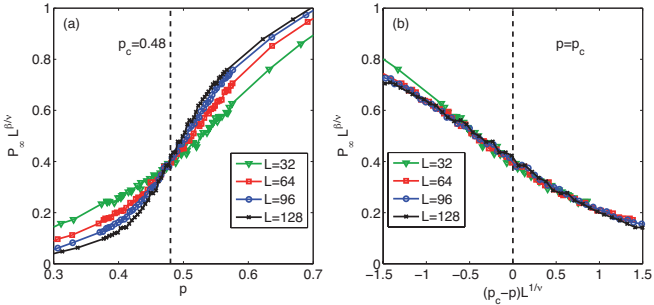}
\end{center}
\caption{(a) Rescaled cluster strength~$P_\infty$ dependence on the filling factor $p$. The ratio $\beta/\nu=0.08\pm0.01$ yields the same value for~$p_\mathrm{c}$ as its previous estimation using~$\Pi$. (b) Rescaled cluster strength~$P_\infty$ as a function of the rescaled parameter~$(p_\mathrm{c}-p) L^{1/\nu}$, allowing one to reduce the uncertainty on the correlation exponent, $\nu=1.6\pm0.1$.} 
\label{fig:FSS_Pinfty}
\end{figure}

The exponent~$\beta$ defines the cluster strength behaviour as $P_\infty \sim (p-p_\mathrm{c})^{\beta}$ for $p\geq p_\mathrm{c}$. Within the finite-size scaling hypothesis, introducing the proper function~$g_\infty$, the cluster strength $P_\infty$ is given by:
\begin{equation}
P_\infty\left(p\sim p_\mathrm{c},L\right) = L^{-\frac{\beta}{\nu}} g_\infty\left[L^{\frac{1}{\nu}}(p-p_\mathrm{c})\right],
\label{eqn:FSS_Pinf}
\end{equation}
so that plotting $P_\infty L^{\frac{\beta}{\nu}}$ with respect to $p$ (Figure~\ref{fig:FSS_Pinfty}a) should give the same unique crossing point as the percolation cumulant. Indeed, we find $\beta/\nu=0.08\pm0.01$,and $p_\mathrm{c}=0.48\pm0.01$, remarkably consistent with the previous determination. Plotting the rescaled order parameter with respect to $L^{1/\nu}(p_\mathrm{c}-p)$ allows one to refine the previous determination of $\nu=1.6\pm0.1$ by seeking a curve collapse (Figure~\ref{fig:FSS_Pinfty}b), from which we deduce $\beta=0.08\pm0.02$.

\begin{figure}[t]
\begin{center}
\includegraphics[width=1.0\columnwidth]{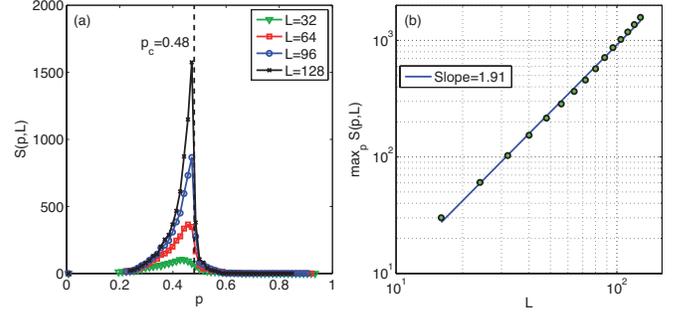}
\end{center}
\caption{(a) Average cluster size~$S$ with respect to $p$. Note that we excluded the percolating cluster from the calculation for $p>p_\mathrm{c}$. (b) Average cluster size with respect to the system size $L$, allowing us to find $\gamma/\nu = 1.91\pm0.05$.} 
\label{fig:FSS_S_L}
\end{figure}

Finally, we determined $\gamma$ relying on the average cluster size. Using the finite-size scaling ansatz, $S$ is expected to increase as $L^{\frac{\gamma}{\nu}}$ close to $p_\mathrm{c}$. Given the value of~$\nu$, we obtain $\gamma=3.0\pm0.3$ (Figure~\ref{fig:FSS_S_L}).

\begin{figure}[t]
\begin{center}
  \includegraphics[width=1.00\columnwidth]{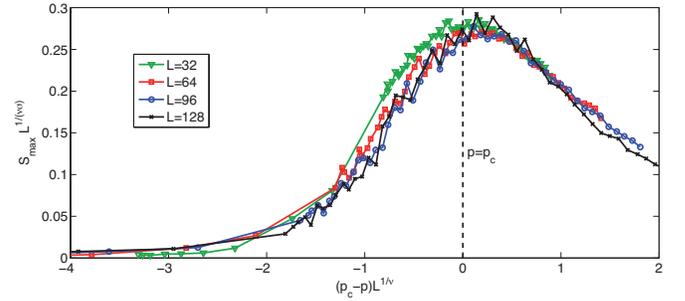}
\end{center}
\caption{Rescaled largest non-percolating cluster as a function of the rescaled parameter~$(p_\mathrm{c}-p) L^{1/\nu}$. Seeking a curve collapse leads to the determination of $\sigma=0.33\pm0.02$.} 
\label{fig:FSS_sigma}
\end{figure}

Finally, the exponent $\sigma$ characterizes the mass divergence of the largest non-percolating cluster when approaching criticality, namely $s_\mathrm{max}\sim|p-p_\mathrm{c}|^{-\sigma}$. Finite-size scaling ansatz therefore implies that
\begin{equation}
s_\mathrm{max}\left(p\sim p_\mathrm{c},L\right)=L^{\frac{1}{\nu\sigma}} g_\mathrm{M} \left[L^{\frac{1}{\nu}}\left(p_\mathrm{c}-p\right)\right],
\end{equation}
where, again, $g_\mathrm{M}$ is a scaling function. Using $\nu=1.6$, we obtain a data collapse for $\sigma=0.33\pm0.02$, as displayed on~Figure~\ref{fig:FSS_sigma}.

The estimation of the exponents~$\alpha$, $\delta$ and~$\nu$ can be done using the hyperscaling relations $\alpha=2-d \nu$ (Josephson's identity), $\gamma=(2-\eta)\nu$, and $\delta-1=\gamma/\beta$ (Widom's identity). They lead to $\alpha=-1.2\pm0.2$, $\delta=24.9\pm6.5$ and~$\eta=0.09\pm 0.27$. Note that the value of $\alpha$ is consistent with Rushbrooke's hyperscaling relation $\alpha+2\beta+\gamma=2$.

\bibliography{PercoBib}

\end{document}